\documentclass[conference]{IEEEtran}
\IEEEoverridecommandlockouts
\usepackage{cite}
\usepackage{amsmath,amssymb,amsfonts}
\usepackage{algorithmic}
\usepackage{graphicx}
\usepackage{textcomp}
\usepackage{xcolor}
\def\BibTeX{{\rm B\kern-.05em{\sc i\kern-.025em b}\kern-.08em
    T\kern-.1667em\lower.7ex\hbox{E}\kern-.125emX}}
\begin{document}

\title{GCS-ICHNet: Assessment of Intracerebral Hemorrhage Prognosis  using Self-Attention with Domain Knowledge Integration \\
\thanks{This work was supported by the Zhejiang Provincial Natural Science Foundation of China (No.LY21F020017, 2022C03043), 
National Natural Science Foundation of China (U20A20386), 
GuangDong Basic and Applied Basic Research Foundation (2022A1515110570), Chinese Key-Area Research and Development Program of Guangdong Province (2020B0101350001), 
Innovation teams of youth innovation in science and technology of high education institutions of Shandong province (2021KJ088), the Shenzhen Science and Technology Program (JCYJ20220818103001002). \\
$\dagger Equal \hspace{0.1cm} Contribution$, $\star Corresponding \hspace{0.1cm} Author$\\}
}
\author{Xuhao Shan$^{1,\dagger}$,  Xinyang Li$^{2,\dagger}$, Ruiquan Ge$^{1,\star}$, Shibin Wu$^3$, Ahmed Elazab$^4$, Jichao Zhu$^5$, Lingyan Zhang$^5$,\\
Gangyong Jia$^{1}$, Qingying Xiao$^{6}$, Xiang Wan$^{7}$, Changmiao Wang$^{7,\star}$\\
 
$^1$Hangzhou Dianzi University, $^2$The Chinese University of Hong Kong, Shenzhen,\\ 
$^3$Ping An Technology,$^4$Shenzhen Universixty,$^5$Longgang District Central Hospital of Shenzhen,\\
$^6$National Health Data Institute, Shenzhen, $^7$Shenzhen Research Institute of Big Data\\

\tt20060309@hdu.edu.cn, \tt120090345@link.cuhk.edu.cn, \\
\tt gespring@hdu.edu.cn, \tt wushibin37@163.com, \tt ahmedelazab@szu.edu.cn,\\
\tt jackeyzjc@163.com, \tt18819818005@163.com, \tt gangyong@hdu.edu.cn,\\
\tt571030498xiao@gmail.com,\tt wanxiang@sribd.cn, \tt cmwangalbert@gmail.com\\
}

\maketitle

\begin{abstract}
Intracerebral Hemorrhage (ICH) is a severe condition resulting from damaged brain blood vessel ruptures, often leading to complications and fatalities. Timely and accurate prognosis and management are essential due to its high mortality rate. However, conventional methods heavily rely on subjective clinician expertise, which can lead to inaccurate diagnoses and delays in treatment. Artificial intelligence (AI) models have been explored to assist clinicians, but many prior studies focused on model modification without considering domain knowledge.
This paper introduces a novel deep learning algorithm, GCS-ICHNet, which integrates multimodal brain CT image data and the Glasgow Coma Scale (GCS) score to improve ICH prognosis. The algorithm utilizes a transformer-based fusion module for assessment. GCS-ICHNet demonstrates high sensitivity  {\bfseries 81.03\%}  and specificity {\bfseries 91.59\%}, outperforming average clinicians and other state-of-the-art methods. The code is available at https://github.com/Windbelll/Prognosis-analysis-of-cerebral-hemorrhage.

\end{abstract}


\section{Introduction}
Intracerebral Hemorrhage (ICH) occurs when blood vessels rupture within the brain parenchyma, leading to bleeding in the brain tissue. As the second most common type of stroke, ICH is associated with high fatality rates \cite{hong2021association}. The global annual incidence rate of ICH ranges from 24.6 to 52.4 cases per 100,000 people, with China experiencing an approximate rate of 69.6 cases per 100,000 people \cite{poon2016epidemiology,van2010incidence}. Unfortunately, the mortality rate within 30 days of ICH remains considerably high, varying from 35\% to 52\% \cite{jakubovic2012intracerebral}. Furthermore, a significant percentage of patients, between 61\% and 88\%, are likely to experience long-term loss of functional independence, which imposes a substantial burden on both society and families \cite{van2010incidence}. Given the high morbidity, disability, and mortality associated with ICH, accurate diagnosis and treatment are essential for improving patients' chances of survival. Clinicians typically evaluate the prognosis of ICH based on initial bleeding volume and hematoma location, which helps in formulating an appropriate treatment plan \cite{kotruchin2022impact}. Key indicators in the diagnosis of ICH include hemorrhage volume, the Glasgow Coma Scale (GCS) score, presence or absence of ICH, and other factors \cite{gurevitz2022association, TEASDALE197481}. Moreover, prognostic outcomes are determined using the Glasgow Outcome Scale (GOS), a rating scale that assesses patients' functional outcomes following brain injury. 
The scoring criteria for GOS are listed in Table \ref{tab:GOS}. \par

\begin{table}
\caption{GOS scoring criteria represent the degree of traumatic brain injury, with scores ranging from 1, indicating the worst outcome, to 5, signifying the best outcome. A GOS score above 4 is considered indicative of a good prognosis.}\label{tab:GOS}
\centering
\resizebox{.45\textwidth}{!}{
\begin{tabular}{|c|c|}
\hline
Score & Categories\\
\hline
1 & Dead\\
 2 & Vegetative state: unconscious and unaware of their surroundings\\
 3 & Severe disability: conscious but dependent on others for daily living activities\\
 4 & Moderate disability: independent but has some physical or cognitive deficits\\
 5 & Good recovery: pre-injury level of the function\\
\hline
\end{tabular}
}
\end{table}
Computed tomography (CT) is widely utilized in ICH diagnosis due to its imaging reconstruction techniques and rapid processing capabilities \cite{huisman2005intracranial}. In clinical settings, timely and accurate assessment of severe and acute ICH is critical when an ICH event occurs. GCS consists of three components: eye-opening, verbal response, and motor response, each with a score that contributes to a total score between 3 (deep coma) and 15 (normal consciousness)\cite{middleton2012practical}. It helps medical professionals assess and communicate the severity of a patient's condition, particularly in cases of head injuries or altered consciousness. Traditional methods involve using CT to determine the GCS score, which is then used to predict the patient's GOS score. However, CT interpretation requires qualified clinicians and this process can be influenced by their experience and subjective judgment. Furthermore, ICH carries a poor prognosis, with only 20\% of survivors fully recovering six months after the incident \cite{hostettler2019intracerebral}. AI has the potential to help overcome these limitations.\par
 AI methods have demonstrated promising results in medical imaging interpretation, such as the diagnosis of breast and prostate cancer \cite{cirecsan2013mitosis,litjens2016deep}, achieving accuracy levels equivalent to or even surpassing those of specialist physicians \cite{emblem2015generic,warman2020interpretable}. 
 However, the application of AI techniques in the field of ICH has primarily focused on the identification and volume measurement of hemorrhage \cite{arbabshirani2018advanced,kim2022cerebral,scherer2016development}. These studies primarily emphasize accuracy at the scan level, rather than at the more granular slice level, which is a key aspect addressed in our work.
 Some studies have targeted improving quality metrics and classification \cite{davis2022machine,zhu2021machine}. Nonetheless, few studies are available for ICH prognosis prediction. For instance, the random forest was employed based on the ICH score and achieved only 76\% prognosis accuracy, which could lead to incorrect or misleading treatment decisions \cite{nawabi2021imaging}. Previous studies have concentrated on model modification rather than the data itself. Incorporating the expertise of skilled radiologists and domain knowledge, such as GCS scores, could result in more accurate prognosis predictions. Therefore, this study aims to establish a prognosis prediction model for ICH patients based on CT images and GCS scores, which may facilitate rapid assessment of ICH severity and provide a basis for making individualized clinical treatment decisions.

\section{Method}

\subsection{Motivation}
Incorporating the GCS score as domain knowledge is motivated by the following reasons: (1) The GCS score serves as an objective assessment of consciousness in ICH patients, as determined by professional physicians. This factor can aid the network training in understanding ICH-related characteristics; (2) The GCS score can provide additional input for classification. However, integrating domain knowledge into a network for better patient prognosis prediction is a challenging task.

To overcome this challenge, we introduce the GCS-ICHNet. This network combines GCS scores and ICH data through separate feature extraction modules for each modality. The features extracted from both modalities are then fused in a dedicated module, allowing us to capture the interplay between the modalities. Detailed explanations of each component are provided in the subsequent subsections.
\subsection{Problem Formalization}\label{s2-2}
Let $\mathbf{P}$ be a dataset consisting of information from $\mathrm{N}$ patients, denoted as $\mathbf{P} = {(\chi_{i},g_{i},\gamma_{i})}_{i=1}^{\mathrm{N}}$, where $\chi_{i}$ represents one or more CT slices associated with the bleeding area, $g_{i}$ is the GCS score of patient $i$, and $\gamma_{i}$ is the prognostic outcome given by clinicians. Our objective is to develop a model $\mathbf{h}$ that approximates the ground truth outcome $\gamma_{i}$ given the input $(\chi_{i},g_{i})$, i.e., $\gamma_{i} \approx \mathbf{h}(\chi_{i},g_{i})$. During the training step, the test labels are masked.
\begin{figure*}
\begin{center}
\includegraphics[width=0.8\textwidth]{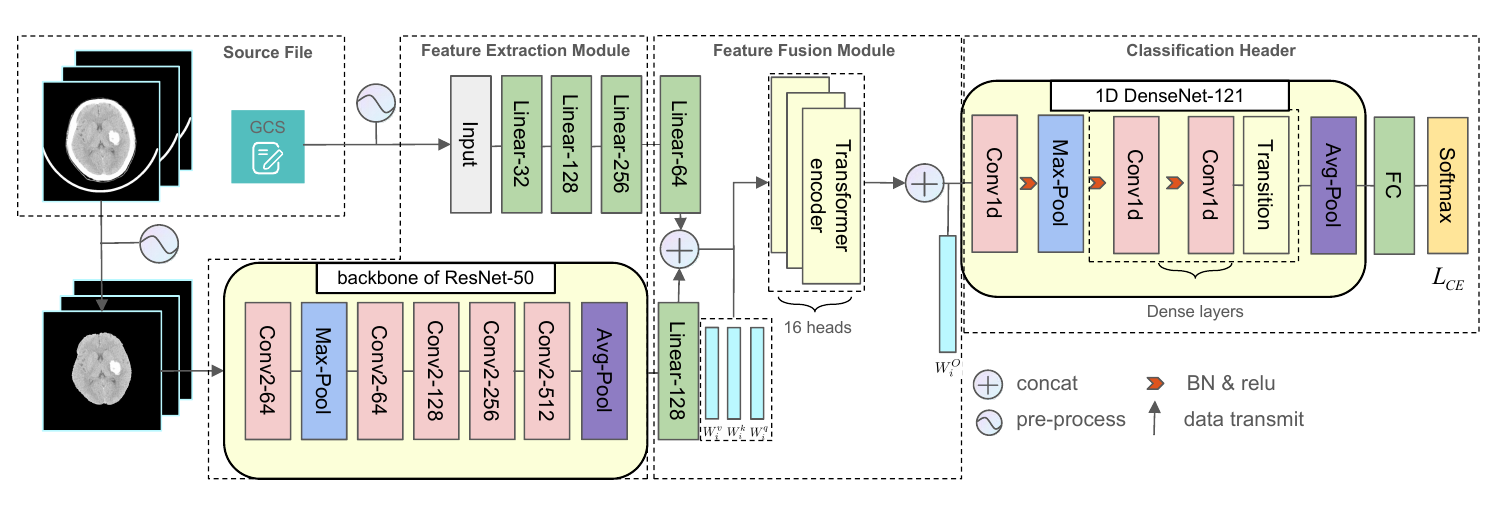}
%\fbox{\rule{0pt}{2in} \rule{.9\linewidth}{0pt} }
\end{center}
   \caption{Architecture of the proposed GCS-ICHNet. The GCS score is encoded by a linear layer, and the CT slice features are extracted through convolution. The features are then fused by a transformer encoder and fed into the classification heads to obtain the final result.}
\label{arch}
\end{figure*}
\subsection{GCS-ICHNet Architecture}\label{s2-3}
The architecture of the proposed GCS-ICHNet is depicted in Figure \ref{arch}. It is composed of three primary components: the feature extraction module, the feature fusion module, and the final classification. We employ the ResNet-50 backbone in the feature extraction module to extract features from the CT slices. Utilizing ResNet-50 also helps to mitigate potential overfitting issues caused by insufficient data. Subsequent to the trunk portion, we incorporate multiple linear layers and Dropout layers to acquire the ultimate CT slice features, which form a one-dimensional vector with a length of 128. To integrate domain knowledge, specifically the GCS score, we utilize a Multilayer Perceptron (MLP) to extract the GCS score and then connect its textual features to the feature extraction module. The MLP consists of three hidden layers with 32, 128, and 256 neurons, and ultimately generates a one-dimensional vector with a length of 64. The GCS score and the features extracted from CT slices are adjusted to the same dimension to facilitate the mixing process. In the feature fusion module, we initially concatenate the image features acquired from CT images and the GCS score features using:
\begin{equation}
f_{global} = unsqueeze(Concat(Conv(\chi_{i}),w{g_{i}}+b)),
\end{equation}
where $f_{global}$ represents the features connected by CT and GCS, $w$ and $b$ are the weights and biases of the linear layer, and $i$ denotes the index of the dataset. The unsqueeze function is employed to increase their dimensionality, enabling them to pass through the subsequent encoder. The mixed feature vector is then passed through a transformer encoder, which is based on a self-attention module comprised of 16 attention heads, to obtain the fused features of both modalities. By utilizing the self-attention mechanism~\cite{vaswani2017attention}, we can more effectively capture the potential relationship between the GCS score and CT images. This can be mathematically expressed as follows:
\begin{equation}
f_{mix} = Concat(head_1,...,head_{j})W^O ,
\end{equation}
meanwhile,
\begin{equation}
head_j = softmax(\frac{Q_{j}K_{j}^{T}}{\sqrt{d_{k}}})V_{j}, j \in (1, 16),
\end{equation}
where $f_{mix}$ represents the mixed feature, $W^O$ is the output weight matrix, $Q_{j},K_{j},V_{j}$ are the results obtained by multiplying the weight matrices in $f_{global}$ and different attention heads, while $d_{k}$ and $j$ are the dimensions of the vector key and the index of different attention heads, respectively. Finally, the fused feature is passed to the classification head 1D DenseNet-121 to overcome overfitting and ensure effective generalization. During the training process, Adam~\cite{adam2014method} is employed as an optimizer and the cross-entropy function is adopted to calculate the loss of the result.
\section{Experiments}
This study involves a total of 286 patients who were admitted to our collaborative hospital between August 2013 and May 2021 and completed the entire treatment cycle for spontaneous ICH. CT scans were acquired using the Philips Brilliance 16-slice CT scanner and the Toshiba Aquilion ONE 320-slice CT scanner.
\subsection{Data Acquisition and Pretreatment}\label{s3-1}
To meet the inclusion criteria for the study, patients were required to fulfill the following criteria: (1) diagnosis of spontaneous ICH; (2) CT scan within 24 hours after the bleeding; (3) complete information regarding the GCS score upon admission; (4) available information about their prognosis based on GOS score upon hospital discharge; (5) records of clinical information such as age, sex, and bleeding site. The retrospective nature of this study has been approved by the hospital ethics committee, and oral consent was obtained from all patients.\par
To ensure that the training data focuses on the patient's bleeding areas and does not contain excessive invalid information, we screened the images under the guidance of clinicians, since the bleeding area, size, and shape varied among different patients, as demonstrated in the prior study \cite{wang2021deep}. This process ensured that each image contained bleeding areas. The selected data were standardized by calculating the mean and standard deviation. Finally, professional clinicians verified the validity of the dataset.\par
Irrelevant information in the images, such as hyperdense intracranial structures, could potentially introduce bias and mislead the classification results. To mitigate interference from non-brain tissue regions, we employ multiple threshold segmentation and connected component screening to isolate these areas. Specifically, we use the OpenCV library to perform threshold segmentation to separate bone and CT slices, and then retain the bleeding area and normal brain tissue. To eliminate excess noise and ensure the accuracy of the results, we perform connected component screening in batches. The final image should only contain information pertaining to the brain tissue area, including the bleeding area, gray matter, and normal brain tissue.

\subsection{Training process and Evaluation criteria}\label{s3-2}
In our experiments, we assess the performance of the network using the following two datasets:\par
1). The CT slice dataset as a baseline: Each item in $\mathbf{P}_i$ contains only CT slice $\chi_i$, without the accompanying GCS score $g_{i}$.\par
2). GCS-ICHNet dataset: Both CT slice data $\chi_i$ and GCS score $g_{i}$ data.\par
The dataset consists of a total of 286 patients. To ensure that CT slices from the same patient are allocated to the same fold, patients are initially randomly divided into 80\% for training and 20\% for testing. Subsequently, the corresponding CT slices of each patient are assigned to their respective CT slice datasets. We performed multiple segmentations to carry out cross-validation. For instance, in one segmentation, 232 patients were designated for training and 54 for testing. The CT slices dataset comprises 1536 CT slices, with 1267 allocated for training and 269 for testing. The experiment aims to classify individual CT slices and patient accuracy is determined through a voting mechanism. A patient's prediction result is considered correct if more than half of their slices are accurately predicted.\par
\begin{table}[h]
\caption{Comparison of Classification Methods: The Accuracy@patient refers to the accuracy of a single patient and is derived from the mean value of multiple cross-validation experiments. This metric enables a comprehensive understanding of the performance of various classifications.}\label{tab2}
\centering
\resizebox{.45\textwidth}{!}{
\begin{tabular}{|c|c|}
\hline
Method &  Accuracy@patient (\%)\\
\hline
GCS only & 46.25 \\
baseline \cite{he2016deep,huang2017densely} &  76.67\\
{\bfseries GCS-ICHNet (ours)} &   {\bfseries85.70}\\

\hline
\end{tabular}}

\end{table}
\subsection{Classification Results}\label{s3-3}
To evaluate the effectiveness of incorporating domain knowledge, we compared GCS-ICHNet with the baseline method, which solely utilizes the ResNet-50 and DenseNet-121 backbones for classification. Additionally, we investigated the impact of varying sample sizes on patient data during the screening process. To accomplish this, we conducted multiple experiments and analyzed the corresponding results. All experiments were performed on two Nvidia RTX 3090Ti GPUs, with a training epoch of 300, a learning rate of 0.0001, and a batch size of 32.\par

While no universally recognized GCS threshold exists to predict a definitive prognosis, clinical doctors commonly consider a GCS score of 9 or higher as indicative of mild head injury, associated with a better prognosis for patients. The findings from Table \ref{tab2} indicate that our proposed GCS scoring method significantly enhances individual patient performance. Specifically, our method displays a remarkable 39.45\% increase compared to the GCS score alone. Furthermore, there is a notable improvement of 9.03\% when compared to the baseline method. These results suggest that our proposed method has the potential to provide more precise prognostic assessments for patients with traumatic brain injuries. However, additional studies with larger sample sizes and multi-center trials are necessary to confirm the clinical utility of our method. Ultimately, the objective of using a more accurate prognostic assessment is to enhance patient outcomes and reduce the burden of traumatic brain injury on individuals, families, and society.\par
\begin{figure}[h]
\center
\includegraphics[width=0.4\textwidth]{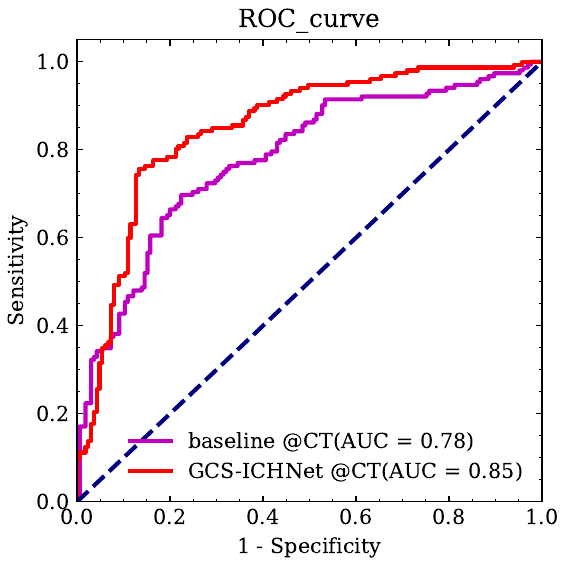}
\caption{ROC curves for the baseline and GCS-ICHNet.} \label{fig2}
\end{figure}
As demonstrated in Figure \ref{fig2}, solely utilizing CT images generates satisfactory classification results. However, integrating the GCS score increases the area under the curve (AUC) value by 7\%. This result indicates that the GCS score significantly enhances the diagnostic efficiency of GCS-ICHNet.

\begin{figure}[h]
\center
\includegraphics[width=0.5\textwidth]{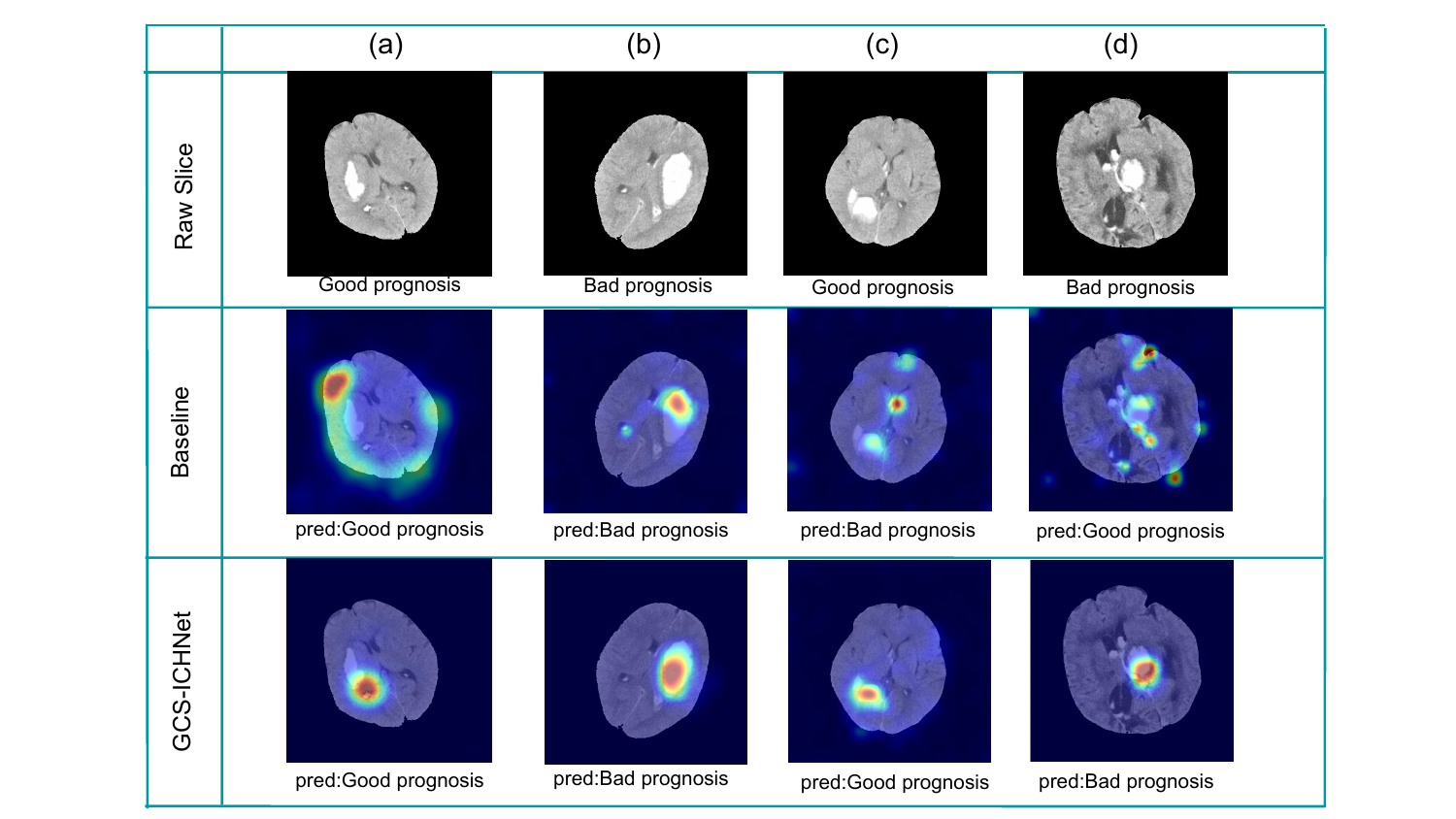}
\caption{The figure depicts visual examples of various prediction results. The class association activation area of the network for the prediction results is highlighted in red.} \label{fig3}
\end{figure}
\subsection{Visualization Analysis}\label{s3-4}
In order to gain insights into the challenges of the task and evaluate the network's prediction process, a subset of images from the test dataset was selected for analysis. The Smooth Grad-CAM method~\cite{omeiza2019smooth} was employed to visualize the last convolutional layer, specifically layer 4 of the ResNet-50 backbone, within the feature extraction module applied to the CT slices. Figure \ref{fig3} presents the various outputs obtained from the two models. In Figure \ref{fig3}a and Figure \ref{fig3}b, both models generate accurate predictions. However, in Figure \ref{fig3}c, the regions of interest are emphasized, highlighting their importance in the predictive capabilities of the two methods for identifying patients with favorable prognoses. Notably, GCS-ICHNet demonstrates superior accuracy in precisely pinpointing the cerebral hemorrhage area, leading to more accurate predictions when compared to the baseline method. Moreover, Figure \ref{fig3}d showcases a compelling scenario in which GCS-ICHNet successfully predicts the correct results even in challenging cases involving complex CT textures. This specific example underscores the robustness and effectiveness of GCS-ICHNet in handling intricate scenarios. The visualized outcomes also indicate that the model integrated with the GCS score excels at extracting relevant features from the CT slice, in contrast to the baseline model, which struggles in this regard.

\subsection{GCS-ICHNet vs. Neurosurgeons}\label{s3-5}
\begin{table}[t]
\caption{Classification comparisons on CT slices between GCS-ICHNet and assessments by neurosurgeons. The neurosurgeons who conducted the assessments had 2.5, 3.5, and 14 years of experience, and are referred to as 'Neurosurgeon1', 'Neurosurgeon2', and 'Neurosurgeon3', respectively. Throughout the study, 'CT' denotes the baseline dataset, while 'CT+GCS' refers to the GCS-ICHNet dataset. 
%The sensitivity and specificity of previous results are represented by TPR and TNR, respectively.
}\label{tab3}
\centering
\resizebox{0.5\textwidth}{!}{
\begin{tabular}{|c|c|c|c|c|c|c|}
\hline
Test &  CT (\%)& TPR (\%) & TNR (\%) & CT + GCS (\%) & TPR (\%) & TNR (\%)\\  
\hline
Neurosurgeon1 &  57.75 & 27.40 & 94.48 & 54.75 & 22.79 & 91.89\\
Neurosurgeon2 &  58.00 & 25.11 & {\bfseries97.79} & 67.25 & 41.86 & {\bfseries96.76} \\  
Neurosurgeon3  &  {\bfseries\color{blue}73.25} & 69.41 & 77.90 &{\bfseries\color{blue}77.50} & 73.02 & 82.70 \\
\hline
Neurosurgeon Avg. & {\bfseries\color{red}63.00} & 40.64 & 90.05 & {\bfseries\color{red}66.50} & 45.89 & 90.45 \\
baseline\cite{he2016deep,huang2017densely} & {\bfseries77.75} & {\bfseries78.54} & 76.80 & - & - & - \\
{\bfseries GCS-ICHNet (ours)} &  - & - & -  & {\bfseries88.00} & {\bfseries85.12} & 91.35 \\
\hline
\end{tabular}}

\end{table}
To assess the effectiveness of our method in clinical practice, we conducted a study in which we randomly selected two groups of 400 CT slices from the dataset, representing 223 different patients. These images were encrypted, and all patient-related information was removed. One group consisted solely of the source images, while the other displayed both the source image and the corresponding GCS score. We invited professional neurosurgeons with varying levels of expertise to predict the prognosis of the data and compared their results with those obtained using our method. The results are summarized in Table \ref{tab3}. Our findings suggest that the inclusion of the GCS score as an auxiliary diagnostic tool can improve the accuracy of prognosis prediction by {\bfseries\color{red}3.50\%}  for {\bfseries\color{red}Neurosurgeon Avg.}, and by {\bfseries10.25\%} for {\bfseries GCS-ICHNet} compared to baseline. However, the GCS score has limited utility in improving the accuracy of prognosis prediction for patients with stricter diagnostic criteria. This indicates that neurosurgeons may subjectively discount the significance of the GCS score, despite its potential usefulness as an indicator. Notably, neurosurgeons often exhibit high specificity and low sensitivity in their assessments, which arises from less experienced neurosurgeons often considering all test cases as negative, and their assessments being highly subjective, regardless of whether they rely solely on CT scans or incorporate GCS scores. In contrast, our method consistently delivers more reliable results and showcases exceptional performance.
\begin{table}[h]
\caption{Classification comparisons on individual patients between GCS-ICHNet and assessments by neurosurgeons. The assessments in this study were conducted by neurosurgeons with 2.5, 3.5, and 14 years of experience, and are denoted as 'Neurosurgeon1', 'Neurosurgeon2', and 'Neurosurgeon3', respectively. 
%The accuracy of the assessments is denoted by 'ACC', while the sensitivity and specificity of previous results are represented by TPR nand TNR.
}\label{tab4}
\centering
\begin{tabular}{|c|c|c|c|c|}
\hline
Test &  ACC (\%)& TPR (\%) & TNR (\%) & F1-Score\\  
\hline
Neurosurgeon1 &  53.81 & 19.83 & 90.65 & 0.322\\
Neurosurgeon2 &  63.23 & 32.76 & {\bfseries96.26} & 0.481 \\  
Neurosurgeon3  &  {\bfseries\color{blue}72.65} & 65.52 & 80.37 & 0.714 \\
\hline
Neurosurgeon Avg. & {\bfseries\color{red}63.26} & 39.37 & 89.21 & 0.506\\
{\bfseries GCS-ICHNet (ours)} &  {\bfseries86.10} & {\bfseries81.03} & 91.59 & 0.904\\
\hline
\end{tabular}

\end{table}

The results of our study, summarized in Table \ref{tab4}, highlight the performance of different methods in identifying individual patients. Notably, our experimental findings suggest that the accuracy of our proposed {\bfseries GCS-ICHNet} surpasses that of {\bfseries\color{blue}Neurosurgeon3} by a significant margin of {\bfseries\color{cyan}{13.45\%}}. This improvement can be attributed to the incorporation of the GCS score into our GCS-ICHNet model, which enables more precise and accurate predictions.

The superior accuracy achieved by GCS-ICHNet holds significant clinical significance, indicating that our method has the potential to enhance medical professionals' diagnostic abilities and lead to better patient outcomes. By incorporating the GCS score into our model, we can provide clinicians with a valuable tool for prognosis prediction, ultimately contributing to more effective treatment strategies and care for patients with ICH.

\begin{table}[h]
\caption{The classification comparisons between GCS-ICHNet and other methods are presented in terms of 'Accuracy,' which represents the predicting accuracy of patient prognosis results, and 'AUC,' representing the area under the Receiver Operating Characteristic (ROC) curve. %Additionally, 'TPR' refers to sensitivity, while 'TNR' refers to specificity.
}\label{tab5}
\centering
\resizebox{0.5\textwidth}{!}{
\begin{tabular}{|c|c|c|c|c|}
\hline
Methods & Accuracy & AUC & TPR & TNR\\  
\hline
ICH score metrics \cite{gregorio2018prognostic} & 0.74 & 0.80 & 0.74 & 0.76\\
imaging-based classifier \cite{nawabi2021imaging} & 0.72 & 0.80 & 0.73 & 0.71\\
ML-based classifier \cite{wang2019automatic} & 0.77 & 0.84 & 0.77 & 0.76\\
{\bfseries GCS-ICHNet (ours)} & {\bfseries0.85} & {\bfseries0.86} & {\bfseries0.80} & {\bfseries0.83} \\
\hline
\end{tabular}}

\end{table}
\subsection{GCS-ICHNet vs. Other Methods}\label{s3-6}
Prognostic assessment of cerebral hemorrhage often relies on scales like GOS and modified Rankin Scale (mRS), which measure the functional status and daily activities of stroke patients. A lower mRS score (0-6) indicates better function, while a higher score implies more severe impairment, aiding treatment decisions\cite{witsch2021prognostication}. These scales and scoring methods provide key information for treatment decisions.\par
For meaningful comparisons, we treat cerebral hemorrhage prognosis as a binary classification task in our study. We use GOS scores at patient discharge to define outcomes, ensuring comparability with similar studies and maintaining analytical consistency. This rigorous approach strengthens the validity of our findings.\par
Table \ref{tab5} presents three methods: 'ICH score metrics,' 'imaging-based classifier,' and 'Machine Learning (ML)-based classifier.' These methods employ machine learning algorithms. A favorable prognosis is defined as an mRS score of 3 or less, while an unfavorable prognosis corresponds to an mRS score greater than 3.  The ICH score metrics classifier assesses stroke severity and prognosis using clinical and imaging data at admission. The 'imaging-based classifier' relies on image texture features, while the 'ML-based classifier' combines both the ICH score metrics and imaging-based classifiers.\par
Our approach excels in prognosis prediction, achieving a remarkable accuracy improvement of 8\% to 13\% compared to previous methods. Deep neural networks effectively extract valuable information from domain knowledge and CT images. Unlike traditional ML-based methods, they capture complex relationships and patterns, resulting in more precise prognosis predictions. This enhanced accuracy can significantly impact medical decision-making, leading to improved treatment strategies and better patient care.\par
\section{Conclusion}
This paper proposes a novel deep neural network, {\bfseries GCS-ICHNet}, for predicting prognosis by integrating the GCS scores into the processing pipeline. The results demonstrate improved performance by incorporating domain knowledge during processing, leading to better learning of information about the bleeding region. The clinical validation of our model indicates that the integration of the GCS score enables better learning of the bleeding region's information and improves the accuracy of prognosis prediction by an impressive {\bfseries\color{green}22.84\%} in simulated tests, surpassing the average accuracy of {\bfseries\color{red}Neurosurgeon Avg.}. However, we acknowledge the limitations of the current study, particularly the relatively small size of the dataset and image dimension. Therefore, future efforts will focus on collecting a larger dataset and designing a 3-dimensional network model. We will also explore pre-training on medical datasets and utilizing the bleeding area test as an auxiliary tool to improve the model's accuracy further. Additionally, we will gather more clinical information during the detection procedure to enrich our understanding of the bleeding region.
\par

\section*{Data Acknowledgements}

This study was conducted in accordance with the principles of the Declaration of Helsinki. Approval was granted by the Ethics Committee of Longgang Central Hospital of Shenzhen (2023.10.26/No.2023ECPJ077).

% Please number citations consecutively within brackets \cite{b1}. The 
% sentence punctuation follows the bracket \cite{b2}. Refer simply to the reference 
% number, as in \cite{b3}---do not use ``Ref. \cite{b3}'' or ``reference \cite{b3}'' except at 
% the beginning of a sentence: ``Reference \cite{b3} was the first $\ldots$''

% Number footnotes separately in superscripts. Place the actual footnote at 
% the bottom of the column in which it was cited. Do not put footnotes in the 
% abstract or reference list. Use letters for table footnotes.

% Unless there are six authors or more give all authors' names; do not use 
% ``et al.''. Papers that have not been published, even if they have been 
% submitted for publication, should be cited as ``unpublished'' \cite{b4}. Papers 
% that have been accepted for publication should be cited as ``in press'' \cite{b5}. 
% Capitalize only the first word in a paper title, except for proper nouns and 
% element symbols.

% For papers published in translation journals, please give the English 
% citation first, followed by the original foreign-language citation \cite{b6}.
\bibliographystyle{IEEEtran}
\bibliography{IEEEabrv,egbib}

% Generated by IEEEtran.bst, version: 1.14 (2015/08/26)
\begin{thebibliography}{10}
\providecommand{\url}[1]{#1}
\csname url@samestyle\endcsname
\providecommand{\newblock}{\relax}
\providecommand{\bibinfo}[2]{#2}
\providecommand{\BIBentrySTDinterwordspacing}{\spaceskip=0pt\relax}
\providecommand{\BIBentryALTinterwordstretchfactor}{4}
\providecommand{\BIBentryALTinterwordspacing}{\spaceskip=\fontdimen2\font plus
\BIBentryALTinterwordstretchfactor\fontdimen3\font minus \fontdimen4\font\relax}
\providecommand{\BIBforeignlanguage}[2]{{%
\expandafter\ifx\csname l@#1\endcsname\relax
\typeout{** WARNING: IEEEtran.bst: No hyphenation pattern has been}%
\typeout{** loaded for the language `#1'. Using the pattern for}%
\typeout{** the default language instead.}%
\else
\language=\csname l@#1\endcsname
\fi
#2}}
\providecommand{\BIBdecl}{\relax}
\BIBdecl

\bibitem{hong2021association}
Y.~Hong, X.-H. Wang, Y.-T. Xiong, J.~Li, and C.-F. Liu, ``Association between admission serum phosphate level and all-cause mortality among patients with spontaneous intracerebral hemorrhage,'' \emph{Risk Management and Healthcare Policy}, pp. 3739--3746, 2021.

\bibitem{poon2016epidemiology}
M.~T. Poon, S.~M. Bell, and R.~A.-S. Salman, ``Epidemiology of intracerebral haemorrhage,'' \emph{New Insights in Intracerebral Hemorrhage}, vol.~37, pp. 1--12, 2016.

\bibitem{van2010incidence}
C.~J. Van~Asch, M.~J. Luitse, G.~J. Rinkel, I.~van~der Tweel, A.~Algra, and C.~J. Klijn, ``Incidence, case fatality, and functional outcome of intracerebral haemorrhage over time, according to age, sex, and ethnic origin: a systematic review and meta-analysis,'' \emph{The Lancet Neurology}, vol.~9, no.~2, pp. 167--176, 2010.

\bibitem{jakubovic2012intracerebral}
R.~Jakubovic and R.~I. Aviv, ``Intracerebral hemorrhage: toward physiological imaging of hemorrhage risk in acute and chronic bleeding,'' \emph{Frontiers in neurology}, vol.~3, p.~86, 2012.

\bibitem{kotruchin2022impact}
P.~Kotruchin, H.~Kliangsa-Ard, T.~Mitsungnern, S.~Imoun, and K.~Kongbunkiat, ``The impact of blood pressure variation on mortality and symptomatic intracerebral hemorrhage in stroke patients after thrombolytic therapy,'' \emph{European Heart Journal}, vol.~43, no. Supplement\_1, pp. ehab849--121, 2022.

\bibitem{gurevitz2022association}
C.~Gurevitz, E.~Auriel, A.~Elis, and R.~Kornowski, ``The association between low levels of low density lipoprotein cholesterol and intracerebral hemorrhage: cause for concern?'' \emph{Journal of Clinical Medicine}, vol.~11, no.~3, p. 536, 2022.

\bibitem{TEASDALE197481}
\BIBentryALTinterwordspacing
G.~Teasdale and B.~Jennett, ``Assessment of coma and impaired consciousness: A practical scale,'' \emph{The Lancet}, vol. 304, no. 7872, pp. 81--84, 1974, originally published as Volume 2, Issue 7872. [Online]. Available: \url{https://www.sciencedirect.com/science/article/pii/S0140673674916390}
\BIBentrySTDinterwordspacing

\bibitem{huisman2005intracranial}
T.~A. Huisman, ``Intracranial hemorrhage: ultrasound, ct and mri findings,'' \emph{European radiology}, vol.~15, no.~3, pp. 434--440, 2005.

\bibitem{middleton2012practical}
P.~M. Middleton, ``Practical use of the glasgow coma scale; a comprehensive narrative review of gcs methodology,'' \emph{Australasian Emergency Nursing Journal}, vol.~15, no.~3, pp. 170--183, 2012.

\bibitem{hostettler2019intracerebral}
I.~C. Hostettler, D.~J. Seiffge, and D.~J. Werring, ``Intracerebral hemorrhage: an update on diagnosis and treatment,'' \emph{Expert review of neurotherapeutics}, vol.~19, no.~7, pp. 679--694, 2019.

\bibitem{cirecsan2013mitosis}
D.~C. Cire{\c{s}}an, A.~Giusti, L.~M. Gambardella, and J.~Schmidhuber, ``Mitosis detection in breast cancer histology images with deep neural networks,'' in \emph{Medical Image Computing and Computer-Assisted Intervention--MICCAI 2013: 16th International Conference, Nagoya, Japan, September 22-26, 2013, Proceedings, Part II 16}.\hskip 1em plus 0.5em minus 0.4em\relax Springer, 2013, pp. 411--418.

\bibitem{litjens2016deep}
G.~Litjens, C.~I. S{\'a}nchez, N.~Timofeeva, M.~Hermsen, I.~Nagtegaal, I.~Kovacs, C.~Hulsbergen-Van De~Kaa, P.~Bult, B.~Van~Ginneken, and J.~Van Der~Laak, ``Deep learning as a tool for increased accuracy and efficiency of histopathological diagnosis,'' \emph{Scientific reports}, vol.~6, no.~1, pp. 1--11, 2016.

\bibitem{emblem2015generic}
K.~E. Emblem, M.~C. Pinho, F.~G. Z{\"o}llner, P.~Due-Tonnessen, J.~K. Hald, L.~R. Schad, T.~R. Meling, O.~Rapalino, and A.~Bjornerud, ``A generic support vector machine model for preoperative glioma survival associations,'' \emph{Radiology}, vol. 275, no.~1, pp. 228--234, 2015.

\bibitem{warman2020interpretable}
A.~Warman, P.~I. Warman, A.~Sharma, P.~Parikh, R.~Warman, N.~Viswanadhan, L.~Chen, S.~Mohapatra, S.~S. Mohapatra, and G.~Sapiro, ``Interpretable artificial intelligence for covid-19 diagnosis from chest ct reveals specificity of ground-glass opacities,'' \emph{medRxiv}, pp. 2020--05, 2020.

\bibitem{arbabshirani2018advanced}
M.~R. Arbabshirani, B.~K. Fornwalt, G.~J. Mongelluzzo, J.~D. Suever, B.~D. Geise, A.~A. Patel, and G.~J. Moore, ``Advanced machine learning in action: identification of intracranial hemorrhage on computed tomography scans of the head with clinical workflow integration,'' \emph{NPJ digital medicine}, vol.~1, no.~1, p.~9, 2018.

\bibitem{kim2022cerebral}
J.-H. Kim, M.~A. Al-masni, S.~Lee, H.~Lee, and D.-H. Kim, ``Cerebral microbleeds detection using a 3d feature fused region proposal network with hard sample prototype learning,'' in \emph{Medical Image Computing and Computer Assisted Intervention--MICCAI 2022: 25th International Conference, Singapore, September 18--22, 2022, Proceedings, Part I}.\hskip 1em plus 0.5em minus 0.4em\relax Springer, 2022, pp. 452--460.

\bibitem{scherer2016development}
M.~Scherer, J.~Cordes, A.~Younsi, Y.-A. Sahin, M.~G{\"o}tz, M.~M{\"o}hlenbruch, C.~Stock, J.~B{\"o}sel, A.~Unterberg, K.~Maier-Hein \emph{et~al.}, ``Development and validation of an automatic segmentation algorithm for quantification of intracerebral hemorrhage,'' \emph{Stroke}, vol.~47, no.~11, pp. 2776--2782, 2016.

\bibitem{davis2022machine}
M.~A. Davis, B.~Rao, P.~A. Cedeno, A.~Saha, and V.~M. Zohrabian, ``Machine learning and improved quality metrics in acute intracranial hemorrhage by noncontrast computed tomography,'' \emph{Current Problems in Diagnostic Radiology}, vol.~51, no.~4, pp. 556--561, 2022.

\bibitem{zhu2021machine}
F.~Zhu, Z.~Pan, Y.~Tang, P.~Fu, S.~Cheng, W.~Hou, Q.~Zhang, H.~Huang, and Y.~Sun, ``Machine learning models predict coagulopathy in spontaneous intracerebral hemorrhage patients in er,'' \emph{CNS Neuroscience \& Therapeutics}, vol.~27, no.~1, pp. 92--100, 2021.

\bibitem{nawabi2021imaging}
J.~Nawabi, H.~Kniep, S.~Elsayed, C.~Friedrich, P.~Sporns, T.~Rusche, M.~B{\"o}hmer, A.~Morotti, F.~Schlunk, L.~D{\"u}hrsen \emph{et~al.}, ``Imaging-based outcome prediction of acute intracerebral hemorrhage,'' \emph{Translational Stroke Research}, vol.~12, pp. 958--967, 2021.

\bibitem{vaswani2017attention}
A.~Vaswani, N.~Shazeer, N.~Parmar, J.~Uszkoreit, L.~Jones, A.~N. Gomez, {\L}.~Kaiser, and I.~Polosukhin, ``Attention is all you need,'' \emph{Advances in neural information processing systems}, vol.~30, 2017.

\bibitem{adam2014method}
K.~D. B.~J. Adam \emph{et~al.}, ``A method for stochastic optimization,'' \emph{arXiv preprint arXiv:1412.6980}, vol. 1412, 2014.

\bibitem{wang2021deep}
X.~Wang, T.~Shen, S.~Yang, J.~Lan, Y.~Xu, M.~Wang, J.~Zhang, and X.~Han, ``A deep learning algorithm for automatic detection and classification of acute intracranial hemorrhages in head ct scans,'' \emph{NeuroImage: Clinical}, vol.~32, p. 102785, 2021.

\bibitem{he2016deep}
K.~He, X.~Zhang, S.~Ren, and J.~Sun, ``Deep residual learning for image recognition,'' in \emph{Proceedings of the IEEE conference on computer vision and pattern recognition}, 2016, pp. 770--778.

\bibitem{huang2017densely}
G.~Huang, Z.~Liu, L.~Van Der~Maaten, and K.~Q. Weinberger, ``Densely connected convolutional networks,'' in \emph{Proceedings of the IEEE conference on computer vision and pattern recognition}, 2017, pp. 4700--4708.

\bibitem{omeiza2019smooth}
D.~Omeiza, S.~Speakman, C.~Cintas, and K.~Weldermariam, ``Smooth grad-cam++: An enhanced inference level visualization technique for deep convolutional neural network models,'' \emph{arXiv preprint arXiv:1908.01224}, 2019.

\bibitem{gregorio2018prognostic}
T.~Greg{\'o}rio, S.~Pipa, P.~Cavaleiro, G.~Atan{\'a}sio, I.~Albuquerque, P.~C. Chaves, and L.~Azevedo, ``Prognostic models for intracerebral hemorrhage: systematic review and meta-analysis,'' \emph{BMC Medical Research Methodology}, vol.~18, pp. 1--17, 2018.

\bibitem{wang2019automatic}
H.-L. Wang, W.-Y. Hsu, M.-H. Lee, H.-H. Weng, S.-W. Chang, J.-T. Yang, and Y.-H. Tsai, ``Automatic machine-learning-based outcome prediction in patients with primary intracerebral hemorrhage,'' \emph{Frontiers in neurology}, vol.~10, p. 910, 2019.

\bibitem{witsch2021prognostication}
J.~Witsch, B.~Siegerink, C.~H. Nolte, M.~Spr{\"u}gel, T.~Steiner, M.~Endres, and H.~B. Huttner, ``Prognostication after intracerebral hemorrhage: a review,'' \emph{Neurological Research and Practice}, vol.~3, pp. 1--14, 2021.

\end{thebibliography}

\end{document}